# PRAGMETIC COMPARISION ANALYSIS OF ALTERNATIVE OPTION PRICING MODELS

Natasha Latif[1], Shafqat Ali Shad.[2]*, Muhammad Usman[4], Chandan Kumar[3], Bahman B Motii[2], MD Mahfuzer Rahman[2], Khuram Shafi[1], Zahra Idrees[1]

COMSATS University Islamabad, Pakistan[1]
University of Montevallo, Montevallo, Alabama, USA[2]
Iowa State University, Iowa, USA[3]
Verizon, Florida, USA[4]
Email: Natasha.latif@outlook.com, sshad@montevallo.edu, chandan@iastate.edu, m.usman30@outlook.com, motiibb@montevallo.edu, mrahman@montevallo.edu, drkhuramshafi@ciitwah.edu.pk, zahra.idrees.butt@gmail.com

**ABSTRACT**- In this paper, we price European Call three different option pricing models, where the volatility is dynamically changing i.e. non-constant. In stochastic volatility (SV) models for option pricing a closed form approximation technique is used, indicating that these models are computationally efficient and have the same level of performance as existing ones. We show that the calibration of SV models, such as Heston model and the High Order Moment based Stochastic Volatility (MSV) is often faster and easier. On 15 different datasets of index options, we show that models which incorporates stochastic volatility achieves accuracy comparable with the existing models. Further, we compare the In Sample and Out Sample pricing errors of each model on each date. Lastly, the pricing of models is compared among three different market to check model performance in different markets.

## 1. INTRODUCTION

The latest summary publication from the world's largest options market, CBOE, reveals a cumulative dollar volume exceeding half a trillion dollars for options in 2014, with over $101 billion traded in equities and more than 477 million transactions recorded (CBOE, 2014). Index options markets have garnered significant attention due to their interaction with stock markets, particularly concerning direction and volume as by Murara (1) and Chance et al. (2).

While American-style exercise prevails in index options, the constant volatility assumption in the standard Black-Scholes model has been empirically challenged. The literature on stochastic volatility, emphasizing aspects like forecast accuracy and option pricing, is extensive and continually expanding as stated by Christoffersen et al. (3). This study aims to model volatility using the MSV model, comparing its accuracy and calibration ease with the Heston model. We categorize volatility models into constant and stochastic, with the latter providing more flexibility in modeling volatility surface as stated by Dumas et al (4), Dupire (5) and Alexander (6).

Our research introduces a new method to model implied volatility, and we focus on establishing its accuracy and ease of calibration, comparing it to existing models. We present heuristic procedures to simplify calibration, with a specific focus on index option pricing. The rest of the paper outlines the two main stochastic volatility models, presents the literature review, empirical methodology, results, and conclusions.

## 2. LITERATURE REVIEW

To address these issues, several studies have proposed extensions to the Black & Scholes model. Two main approaches emerged: deterministic volatility models, which link volatility to observable market variables, and stochastic volatility models, where volatility itself is treated as a stochastic process. This thesis aims to fill two gaps. Firstly, we explore improvements over Black & Scholes by considering both deterministic and stochastic volatility models for pricing OMXS30 index options. The study compares various option pricing models, including Practitioner Black-Scholes, Gram-Charlier, Heston, and Heston Nandi GARCH, using Excel VBA and SAS Enterprise. The dataset consists of call options on OMXS30 from 1st June 2011 to 31st May 2012, focusing on a smaller market like the Swedish Stock Exchange that has received less attention in the existing literature.

Secondly, while there are studies comparing the incremental contribution of stochastic volatility models or more sophisticated models like jump diffusion models, there is a lack of research that systematically compares alternative groups of option pricing models. This thesis contributes as an empirical study that fills this gap by comparing and evaluating various alternative option pricing models sourced by Stein & Stein (7) ; Romo (8) and Schofield (10).

## 3. EMPIRICAL METHODOLOGY

In all the three valuation models, we assume that volatility is a stochastic function of underlying strike price and time to maturity.

### 3.1. Black-Scholes Model

For the price of a non-dividend paying European call option, the Black-Scholes equation is described as Black & Scholes (10) and Cohen (11);

$$C(S,t) = SN(d_1) - Ke^{-r(T-t)}N(d_2) \quad (1)$$

Where $C(S,t)$ is the call option price, S is the stock price at time t, $N(d)$ are the value of cumulative normal distribution, K is the strike price, r is the interest rate, t is the time to maturity, $(T-t)$ is the option duration to get expiry and the $\sigma^2$ is the volatility. Under BS framework the discounted expected value is given by

$$d_1 = \frac{\ln\left(\frac{S}{K}\right) + \left(r + \frac{\sigma^2}{2}\right)(T-t)}{\sigma\sqrt{(T-t)}} \quad (2)$$

$$d_2 = d_1 - \sigma\sqrt{(T-t)} \qquad (3)$$

When the Black-Scholes equation was published, it assumes that during the option life no dividend are paid with no taxes and transaction cost. Also, the risk-free rate is same for all maturities and the interest rate is constant as well as the short selling and trading in continuous time is possible. And the most important of all assumptions is that the volatility of the stock return volatility is constant. Later on this constant volatility framework was modified by Heston by including the stochastic volatility in pricing options.

### 3.2. Heston Stochastic Volatility Model

Heston model assumes that the process $S_t$ follows a log normal distribution, and the process $V_t$ follows a Cox, Cox et al.[12]. For Heston model, the asset price dynamic is assumed to be governed by

$$dS_t = \mu S_t dt + \sqrt{V_t} S_t dW_1(t) \qquad (4)$$

$$dV_t = \chi[\theta - v(t)]dt + \sigma\sqrt{v(t)}dW_2(t) \qquad (5)$$

$$dW_1(t)dW_2(t) = \rho dt \qquad (6)$$

Where μ is the rate of the return of the asset (drift coefficient), $dW_t$ and $dZ_t$ are the standard wiener process with a give correlation coefficients $W_1(t)$ and $W_2(t) = \rho$, and the $\rho, \theta, \sigma, \chi, S_t$ and $V_t$ are the known constants. The price of the European call option with strike price K is given by

$$C_{EUR}(S_t, V_t, t, T) = S_t P_1 - Ke^{-r(T-t)} P_2 \qquad (7)$$

Where the first term is the present value of the spot asset upon optimal exercise and second term is the present value of the strike payment. $S_t$ is the spot price at time t, T is the expiration time and $P_1$ and $P_2$ are the pseudo probabilities. Both P1 and P2 ought to satisfy the PDE.

$$P_j(x, v, \tau, \varphi_j) = \frac{1}{2} + \frac{1}{\pi}\int_0^\infty Re\left[\frac{e^{-i\varphi \ln K} f_j(x, v, \tau, \varphi)}{i\varphi}\right]d\varphi, \quad j = 1,2 \qquad (8)$$

Here, $\tau = T - t$ and $\varphi_j = (v_t, \tau, \chi) = \exp[C_j(\tau, x)\bar{v} + D_j(\tau, x)v_t]$ is the characteristic function, which assumes the characteristic function solution as,

$$C_j(\tau; \varphi_j) = \mu\varphi_j\tau + \frac{a}{\sigma^2}\left\{(b_j - \rho\sigma\varphi_j i + d_j)\tau - 2\ln\left[\frac{1 - g_j e^{d_j\tau}}{1 - g_j}\right]\right\}$$

$$D_j(\tau; \varphi) = \frac{b_j - \rho\sigma\varphi_j i + d_j}{b_j - \rho\sigma\varphi_j - d_j}\left[\frac{1 - e^{d_j\tau}}{1 - g_j e^{d_j\tau}}\right]$$

And

$$g_j = \frac{b_j - \rho\sigma\varphi_j i + d}{b_j - \rho\sigma\varphi_j i - d}$$

$$d_j = \sqrt{(\rho\sigma\varphi_j i - b_j)^2 - \sigma^2(2\mu_j\varphi_j - \varphi_j^2)}$$

$$u_1 = \frac{1}{2}, \quad u_2 = -\frac{1}{2}, \quad a = x\varphi$$

$$b_1 = x + \lambda - \rho\sigma, \quad b_2 = x + \lambda$$

### 3.3. High Order Moments based Stochastic Volatility (MSV)

In MSV model, volatility has a term structure modified by a scalar random variable. And one of the simplest framework to introduced a stochastic component in the volatility is to consider the Hull-White type model of the asset price process by Hull et al. [13].

$$dS_t = rS_t dt + \sqrt{v_t} S_t dW_t^1 \qquad (9)$$

$$dV_t = f_1(t, v_t)dt + f_2(t, v_t)dW_t^2 \qquad (10)$$

Where, $W_t^1$ and $W_t^2$ are the uncorrelated Wiener Process and $f_1, f_2$ are smooth functions bounded by liner growth. Let $\bar{V}$ be the mean variance over some time interval $[0,T]$ defined by

$$\bar{V}_t = \frac{1}{T}\int_0^T \sigma^2(t)dt \qquad (11)$$

Whereas the price of European call option at time 0, for a time to maturity τ can be derived as expectation of Black Scholes price with respect to the variance rate:

$$C_{EUR} = \mathbb{E}\left[C_{BS}\frac{1}{\tau}\int_0^T v_t dt\right] \qquad (12)$$

Where $C_{BS}(x)$ denotes Black-Scholes price evaluated at variance x. the above formula is independent of the exact process followed by $v_t$ (under normal assumption about t-continuity and uniqueness). Denoting the Variance rate $\frac{1}{\tau}\int_0^\tau v_t dt$ by $\bar{V}_t$ and assuming that the moments in question exist. And by expanding the right hand side of the above equation around $\mathbb{E}(\bar{V}_\tau)$ in Tylor series as

$$C_{EUR} \approx C_{BS}(\bar{V}_t) + \sum_{i=2}^M \frac{\partial^i C_{BS}}{\partial \bar{V}_t^i} \frac{\mathbb{E}(\bar{V}_t - \mathbb{E}(\bar{V}_t))^i}{i!}$$

Where the partial derivatives can be calculated at $\mathbb{E}(\bar{V}_t)$. Than they construct a process for $v_t$ which make the right hand side of the above equation easy to evaluate. Then they assume that $v_t$ in equation 10 is governed by the following specific stochastic process

$$dv_t = (\mu_t dt + \gamma_t dW_t^2)v_t \qquad (14)$$

Where, $\mu_t$ is the positive deterministic and integral function, $\gamma_t$ is the positive deterministic function which is piecewise continues with $\gamma_t = 0, t > t_0$ and $W_t^2$ is a standard Wiener process uncorrelated with $W_t^1$. Using Ito's lemma, it is straightforward to show

$$v_t = \exp\left(\int_0^t \mu_s ds\right)\xi_t$$

Where $\xi_t$ is a log normal process with unit mean and a constant variance for $t > t_0$. In particular

$$Var(\xi_t) = \left(\exp\left\{\int_0^{t_0} \gamma_s^2 ds\right\} - 1,\right) t > t_0$$

Than they will henceforth assume that $t > t_0$ holds. Let $k = \sqrt{Var(\xi_t)}$ then the third and fourth centered moments of $\xi_t, m_3$ and $m_4$ respectively can be expressed as:

$$m_3 = k^4(3 + k^2) \qquad (15)$$

$$m_4 = k^4\{(1 + k^2)^4 + 2(1 + k^2)^3 + 3(1 + k^2)^2 - 3\} \qquad (16)$$

They will parameterize the standard deviation k of the lognormal random variable $\xi_t$ directly, with no referencw to $\gamma_t$ or $t_0$. Finally they parametrize $\exp(\int_0^t \mu_s ds)$ as

$$\exp\left(\int_0^t \mu_s ds\right) = \widehat{\sigma_0^2} e^{-\lambda t} + \widehat{\sigma_1^2}\lambda t e^{-\lambda t} + \widehat{\sigma_2^2}$$

Where, $\widehat{\sigma_0}, \widehat{\sigma_1}, \widehat{\sigma_2}, \lambda$ are scalar parameter, this gives our variance model parameterization as

$$v_t = \xi_t\left(\widehat{\sigma_0^2} e^{-\lambda t} + \widehat{\sigma_1^2}\lambda t e^{-\lambda t} + \widehat{\sigma_2^2}\right), \qquad (17)$$
$$\xi_t \sim LN(1, k^2), t > t_0$$

This is the MSV model, based on the use of high order moments of the aforementioned random variable (Date & Islyaev, 2015)

## 4. DATA SPECIFICTAION

The sample comprises closing option prices on three stock indices (S&P 500, Volatility Index, and Russell 2000) traded on the Chicago Board Option Exchange (CBOE) across five different days (01 Nov 2012, 26 Nov 2012, 25 Jul 2013, 26 Jul 2013, 29 Jul 2013). Each index has 100 European style options with maturities ranging from 30 days to 1 year and various strike prices. Implied risk-free rates are proxied using the 3-month London Interbank Offered Rate-Overnight Index Swaps (LIBOR-OIS Rates) for each maturity. The calibration and validation of models were performed using option price data obtained from CBOE Live Lol Data shop, resulting in 15 data sets with 100 prices in each set.

## 5. COMPARISION RESULTS

The application of BS and the two stochastic model to the real market data is now discussed. Three different sets of results are used for comparison purpose. For each in-sample and out-of-sample data set after calibration (30 data sets in all with each of 15 data sets split into in-sample and out-of-sample subsets), for comparison we consider two commonly used error metrics, which is Mean Relative Absolute Error (MRAE) and Root Mean Square Error (RMSE).

**Table 1: Optimization Techniques results (MRAE & RMSE)**

| Dates | Error Matrix | SPX BS | SPX Heston | SPX MSV | VIX BS | VIX Heston | VIX MSV | RUT BS | RUT Heston | RUT MSV |
|---|---|---|---|---|---|---|---|---|---|---|
| 7.Mar.2017 | MRAE-I* | 16.18 | 0.94 | 4.65 | 0.49 | 0.38 | 0.20 | 24.85 | 15.63 | 0.51 |
| | RMSE-I* | 275.34 | 62.53 | 8.54 | 9.32 | 2.59 | 0.33 | 4.08 | 4.22 | 11.00 |
| | Time | Nil | 21.09 | 0.28 | Nil | 37.2 | 0.42 | Nil | 23.12 | 0.22 |
| | MRAE-O** | 5.10 | 0.77 | 4.73 | 33.75 | 0.32 | 0.08 | 7.14 | 13.96 | 0.51 |
| | RMSE-O** | 144.84 | 60.97 | 11.40 | 8.62 | 2.51 | 0.27 | 22.29 | 3.56 | 8.74 |
| 3.Mar.2017 | MRAE-I | 130.32 | 5.75 | 4.71 | 45.32 | 8.66 | 4.52 | 4.41 | 4.25 | 4.34 |
| | RMSE-I | 59.57 | 3.50 | 2.99 | 9.30 | 23.58 | 9.35 | 19.50 | 28.49 | 18.35 |
| | Time | Nil | 29.20 | 1.17 | Nil | 667.25 | 0.85 | Nil | 134.27 | 0.98 |
| | MRAE-O | 127.38 | 6.01 | 5.29 | 24.17 | 7.70 | 5.74 | 4.11 | 4.47 | 8.03 |
| | RMSE-O | 59.33 | 3.20 | 2.30 | 8.24 | 14.53 | 10.73 | 19.49 | 19.44 | 18.75 |
| 15.Feb.2017 | MRAE-I | 20.91 | 16.46 | 3.85 | 26.25 | 8.66 | 4.52 | 7.05 | 4.31 | 4.24 |
| | RMSE-I | 5.29 | 2.97 | 3.05 | 12.38 | 23.58 | 9.35 | 25.32 | 34.34 | 17.55 |
| | Time | Nil | 22.80 | 1.7 | Nil | 590.72 | 2.77 | Nil | 286.69 | 0.19 |
| | MRAE-O | 20.71 | 13.20 | 0.28 | 13.90 | 7.70 | 5.74 | 13.55 | 4.19 | 8.01 |
| | RMSE-O | 5.25 | 3.43 | 2.43 | 17.32 | 14.35 | 10.73 | 60.01 | 22.75 | 18.17 |
| 23.Jan.2017 | MRAE-I | 25.58 | 4.76 | 3.97 | 47.64 | 0.34 | 0.10 | 18.10 | 16.11 | 4.51 |
| | RMSE-I | 3.74 | 3.26 | 3.11 | 9.78 | 2.90 | 0.55 | 10.87 | 22.26 | 0.78 |
| | Time | Nil | 252.93 | 4.32 | Nil | 28.68 | 6.8 | Nil | 25.51 | 2.80 |
| | MRAE-O | 5.15 | 6.20 | 0.35 | 22.57 | 0.31 | 0.05 | 4.34 | 10.84 | 5.27 |
| | RMSE-O | 3.74 | 14.31 | 2.34 | 8.53 | 2.86 | 0.01 | 2.16 | 14.31 | 11.11 |
| 30.Dec.2017 | MRAE-I | 226.17 | 38.95 | 0.12 | 0.43 | 0.32 | 0.10 | 238.02 | 126.96 | 0.15 |
| | RMSE-I | 190.09 | 202.01 | 346.62 | 15.82 | 3.73 | 0.58 | 160.66 | 107.39 | 67.07 |
| | Time | Nil | 34.11 | 0.38 | Nil | 28.9 | 0.43 | Nil | 45.02 | 0.34 |
| | MRAE-O | 214.21 | 28.45 | 0.01 | 0.39 | 0.28 | 0.03 | 237.56 | 105.03 | 0.14 |
| | RMSE-O | 189.66 | 198.82 | 10.96 | 15.73 | 3.46 | 0.01 | 160.49 | 92.75 | 67.07 |

*Mean Relative Absolute Error (MRAE): I=In, O=Out
**Root Mean Squared Error (RMSE): I=In, O=Out

This table demonstrate the three different measures of performance In Sample, Out Sample & the computational speed of stochastic volatility models. Overall, MSV performance is really good as compared to other two benchmark models. BS model performance is not good most of it values are worst. In all the three index SPX, VIX & RUT MSV computation speed is very minimum as compared to Heston model. Whereas Heston take much time in calculation because of the complex integrals.

Further, the computational speed is one of the main selling points of any method, we will also compare the stochastic models computational speed it takes for calibrating models parameters. All of the data sets are represented here in **Error! Reference source not found.**, which display the result of In and Out sample in two different error matrices. Row 1 & 2 represents the in sample and row 4 & 5 represents the out sample in both RMSE and MRAE error matrices. Further, row no 3 represents the computational speed of our stochastic volatility model.

The bold face number in each column indicate the worst value of the error matrices. We have total of 180 error value to compare and identified that the Black Scholes model is the one whose overall performance is bad, overall 43 values perform worst in BS model. Than in Heston model total 13 values both in and out sample whose performance is worst and lastly we have total of 2 values in MSV model which tell us that the performance of the MSV model is the best because its computational speed is very low as compare to other stochastic volatility model.

**Table 1: Total number of worst values**

| Models | In Sample | Out Sample |
|---|---|---|
| BS | 22 | 21 |
| Heston | 5 | 8 |
| MSV | 1 | 1 |

**Comparison with Black-Scholes equation**

There are number of measures to check the accuracy of option pricing models. Another measure we used is the pricing error, which is also called the average relative percentage method. The results of price comparison is shown from table 5.3 to 5.5, first four columns in each table represents the real options price, BS price, Heston Price and the MSV price. Then we calculate the error matrices for each model option price. BS price is taken as a benchmark and then we calculate the error between them. After that we apply the dummy on it in which (0 means that stochastic model is better and 1 means that stochastic model is worse). After analyzing all error matrices we come to the conclusion that the stochastic models pricing error matrices perform better in all index. And the stochastic models perform very well as compare to BS model. In the last column, 0 means Stochastic Volatility error is less than BS error, meaning that stochastic model errors is better than BS. And if it shows 1, than it means that BS is superior as compare to stochastic volatility models. There is no doubt that BS error is beaten by heston error in all the three index. But when we compare BS error with MSV error we find different results. In SPX index MSV error is lower than BS error indicating that in SPX index MSV performance is very well but in VIX index we can see that on 30th December and on 23rd January MSV error is beaten by BS error. While in RUT index we can see that only on 7th march MSV error is beaten by BS error but overall we conclude that the Valuation of the MSV model is more accurate. But Heston model also provide accurate results. So we can conclude that if we use stochastic volatility model as a benchmark it may provide us more accurate results because of disturbance parameter included in it and it get adjusted over the time.

**Table 3: Empirical Results of Experiments**

Standard & Poor's Index (SPX)

| Date | Real Price | BS Price | Heston Price | MSV Price | BS Error | Heston Error | Compare | BS Error | MSV Error | Compare |
|---|---|---|---|---|---|---|---|---|---|---|
| 7.March.2017 | 2366.00 | 2362.80 | 2363.32 | 2363.45 | 0.0014 | 0.0011 | 0 | 0.0014 | 0.0011 | 0 |
| | 2366.00 | 2360.99 | 2361.03 | 2363.17 | 0.0021 | 0.0021 | 0 | 0.0021 | 0.0012 | 0 |
| | 2365.35 | 2360.70 | 2357.86 | 2362.12 | 0.0020 | 0.0032 | 0 | 0.0020 | 0.0014 | 0 |
| | 2365.35 | 2359.82 | 2361.73 | 2362.55 | 0.0023 | 0.0015 | 0 | 0.0023 | 0.0012 | 0 |
| | 2365.35 | 2361.83 | 2362.77 | 2363.68 | 0.0015 | 0.0011 | 0 | 0.0015 | 0.0007 | 0 |
| | 2365.35 | 2359.87 | 2360.75 | 2362.01 | 0.0023 | 0.0019 | 0 | 0.0023 | 0.0014 | 0 |
| | 2365.35 | 2360.01 | 2361.86 | 2362.08 | 0.0023 | 0.0015 | 0 | 0.0023 | 0.0014 | 0 |
| | 2366.20 | 2277.71 | 2306.11 | 2306.97 | 0.0374 | 0.0254 | 0 | 0.0374 | 0.0250 | 0 |
| | 2366.20 | 2306.39 | 2306.67 | 2306.83 | 0.0253 | 0.0252 | 0 | 0.0253 | 0.0251 | 0 |
| | 2366.20 | 2364.55 | 2364.72 | 2364.88 | 0.0007 | 0.0006 | 0 | 0.0007 | 0.0006 | 0 |
| 3.March.2017 | 2382.56 | 2377.18 | 2381.69 | 2382.17 | 0.0023 | 0.0004 | 0 | 0.0023 | 0.0002 | 0 |
| | 2382.56 | 2378.33 | 2380.81 | 2382.54 | 0.0018 | 0.0007 | 0 | 0.0018 | 0.0000 | 0 |
| | 2382.56 | 2377.29 | 2381.82 | 2382.03 | 0.0022 | 0.0003 | 0 | 0.0022 | 0.0002 | 0 |
| | 2381.70 | 2315.09 | 2328.49 | 2332.10 | 0.0280 | 0.0223 | 0 | 0.0280 | 0.0208 | 0 |
| | 2381.70 | 2365.11 | 2365.65 | 2367.81 | 0.0070 | 0.0067 | 0 | 0.0070 | 0.0058 | 0 |
| | 2381.80 | 2358.75 | 2364.90 | 2366.39 | 0.0097 | 0.0071 | 0 | 0.0097 | 0.0065 | 0 |
| | 2381.80 | 2360.75 | 2362.48 | 2378.67 | 0.0088 | 0.0081 | 0 | 0.0088 | 0.0013 | 0 |
| | 2381.80 | 2373.46 | 2378.45 | 2379.38 | 0.0035 | 0.0014 | 0 | 0.0035 | 0.0010 | 0 |
| | 2381.80 | 2376.38 | 2378.41 | 2381.35 | 0.0023 | 0.0014 | 0 | 0.0023 | 0.0002 | 0 |
| | 2382.20 | 2378.05 | 2379.19 | 2381.16 | 0.0017 | 0.0013 | 0 | 0.0017 | 0.0004 | 0 |
| 15.February.2017 | 2347.75 | 2308.51 | 2328.07 | 2338.58 | 0.0084 | 0.0167 | 0 | 0.0084 | 0.0039 | 0 |
| | 2347.75 | 2303.86 | 2329.52 | 2338.33 | 0.0187 | 0.0078 | 0 | 0.0187 | 0.0040 | 0 |
| | 2347.75 | 2303.74 | 2328.31 | 2341.17 | 0.0187 | 0.0083 | 0 | 0.0187 | 0.0028 | 0 |
| | 2347.75 | 2304.27 | 2327.98 | 2333.32 | 0.0185 | 0.0084 | 0 | 0.0185 | 0.0061 | 0 |
| | 2347.75 | 2301.64 | 2324.43 | 2336.90 | 0.0196 | 0.0099 | 0 | 0.0196 | 0.0046 | 0 |
| | 2349.51 | 2304.55 | 2331.93 | 2340.53 | 0.0191 | 0.0075 | 0 | 0.0191 | 0.0038 | 0 |
| | 2349.51 | 2301.21 | 2334.24 | 2336.25 | 0.0206 | 0.0065 | 0 | 0.0206 | 0.0056 | 0 |
| | 2349.51 | 2325.96 | 2332.81 | 2342.09 | 0.0100 | 0.0071 | 0 | 0.0100 | 0.0032 | 0 |
| | 2349.93 | 2329.61 | 2334.71 | 2340.69 | 0.0086 | 0.0065 | 0 | 0.0086 | 0.0039 | 0 |
| | 2349.93 | 2329.64 | 2333.62 | 2339.83 | 0.0086 | 0.0069 | 0 | 0.0086 | 0.0043 | 0 |
| 23.January.2017 | 2261.95 | 2246.49 | 2249.33 | 2249.83 | 0.0068 | 0.0056 | 0 | 0.0068 | 0.0054 | 0 |
| | 2259.09 | 2255.09 | 2259.71 | 2260.36 | 0.0013 | 0.0010 | 0 | 0.0013 | 0.0007 | 0 |
| | 2261.95 | 2232.47 | 2241.92 | 2261.57 | 0.0130 | 0.0089 | 0 | 0.0130 | 0.0002 | 0 |
| | 2261.95 | 2233.02 | 2245.22 | 2249.84 | 0.0128 | 0.0074 | 0 | 0.0128 | 0.0054 | 0 |
| | 2262.10 | 2249.16 | 2219.46 | 2237.52 | 0.0057 | 0.0188 | 0 | 0.0057 | 0.0109 | 0 |
| | 2262.10 | 2243.00 | 2247.57 | 2250.09 | 0.0084 | 0.0064 | 0 | 0.0084 | 0.0053 | 0 |
| | 2263.35 | 2230.09 | 2231.44 | 2238.02 | 0.0147 | 0.0141 | 0 | 0.0147 | 0.0112 | 0 |
| | 2263.35 | 2237.75 | 2240.06 | 2252.86 | 0.0113 | 0.0103 | 0 | 0.0113 | 0.0046 | 0 |
| | 2266.68 | 2238.68 | 2255.01 | 2261.73 | 0.0124 | 0.0051 | 0 | 0.0124 | 0.0022 | 0 |
| | 2266.62 | 2257.82 | 2260.80 | 2260.91 | 0.0039 | 0.0026 | 0 | 0.0039 | 0.0025 | 0 |
| 30.December.2017 | 2235.55 | 2212.59 | 2222.61 | 2235.66 | 0.0103 | 0.0058 | 0 | 0.0103 | 0.0000 | 0 |
| | 2235.55 | 2209.68 | 2216.62 | 2224.65 | 0.0116 | 0.0085 | 0 | 0.0116 | 0.0049 | 0 |
| | 2235.55 | 2222.75 | 2227.74 | 2238.76 | 0.0057 | 0.0035 | 0 | 0.0057 | 0.0014 | 0 |
| | 2235.85 | 2219.23 | 2219.74 | 2230.12 | 0.0074 | 0.0072 | 0 | 0.0074 | 0.0026 | 0 |
| | 2235.85 | 2219.92 | 2223.46 | 2226.65 | 0.0071 | 0.0055 | 0 | 0.0071 | 0.0041 | 0 |
| | 2235.85 | 2220.29 | 2221.76 | 2233.29 | 0.0070 | 0.0063 | 0 | 0.0070 | 0.0011 | 0 |
| | 2235.85 | 2220.02 | 2222.01 | 2226.21 | 0.0071 | 0.0062 | 0 | 0.0071 | 0.0043 | 0 |
| | 2235.85 | 2213.12 | 2214.41 | 2223.71 | 0.0102 | 0.0096 | 0 | 0.0102 | 0.0054 | 0 |
| | 2231.60 | 2214.59 | 2222.83 | 2224.61 | 0.0076 | 0.0039 | 0 | 0.0076 | 0.0031 | 0 |
| | 2231.60 | 2220.10 | 2224.87 | 2224.92 | 0.0052 | 0.0030 | 0 | 0.0052 | 0.0030 | 0 |

Volatility Index (VIX)

| Date | Real Price(S) | BS Price | Heston Price | MSV Price | BS Error | Heston Error | Compare | BS Error | MSV Error | Compare |
|---|---|---|---|---|---|---|---|---|---|---|
| 7.March.2017 | 15.35 | 15.03 | 15.18 | 15.19 | 0.0205 | 0.0101 | 0 | 0.0205 | 0.0003 | 0 |
| | 15.35 | 15.07 | 15.18 | 15.24 | 0.0181 | 0.0108 | 0 | 0.0181 | 0.0074 | 0 |
| | 15.35 | 15.18 | 15.28 | 15.35 | 0.0109 | 0.0045 | 0 | 0.0109 | 0.0000 | 0 |
| | 14.38 | 14.08 | 14.27 | 14.33 | 0.0208 | 0.0076 | 0 | 0.0208 | 0.0037 | 0 |
| | 14.38 | 14.10 | 14.27 | 14.37 | 0.0194 | 0.0078 | 0 | 0.0194 | 0.0005 | 0 |
| | 14.38 | 14.20 | 14.35 | 14.32 | 0.0126 | 0.0022 | 0 | 0.0126 | 0.0042 | 0 |
| | 14.38 | 14.19 | 14.34 | 14.35 | 0.0131 | 0.0029 | 0 | 0.0131 | 0.0018 | 0 |
| | 14.38 | 14.12 | 14.18 | 14.31 | 0.0182 | 0.0142 | 0 | 0.0182 | 0.0049 | 0 |
| | 12.48 | 14.10 | 14.17 | 14.30 | 0.1295 | 0.1351 | 0 | 0.1295 | 0.1456 | 0 |
| | 12.48 | 14.20 | 14.23 | 14.37 | 0.1379 | 0.1403 | 0 | 0.1379 | 0.1514 | 0 |
| 3.March.2017 | 15.63 | 15.26 | 15.52 | 15.57 | 0.0239 | 0.0070 | 0 | 0.0239 | 0.0036 | 0 |
| | 14.73 | 14.39 | 14.56 | 14.58 | 0.0230 | 0.0115 | 0 | 0.0230 | 0.0105 | 0 |
| | 14.73 | 14.47 | 14.51 | 14.61 | 0.0173 | 0.0146 | 0 | 0.0173 | 0.0081 | 0 |
| | 14.73 | 14.64 | 14.68 | 14.71 | 0.0062 | 0.0033 | 0 | 0.0062 | 0.0013 | 0 |
| | 14.73 | 14.58 | 14.63 | 14.70 | 0.0103 | 0.0071 | 0 | 0.0103 | 0.0023 | 0 |
| | 12.30 | 12.30 | 12.45 | 12.47 | 0.0408 | 0.0291 | 0 | 0.0408 | 0.0065 | 0 |
| | 12.82 | 12.50 | 12.63 | 12.76 | 0.0253 | 0.0150 | 0 | 0.0253 | 0.0047 | 0 |
| | 12.82 | 12.56 | 12.66 | 12.70 | 0.0035 | 0.0121 | 0 | 0.0035 | 0.0095 | 0 |
| | 12.82 | 12.50 | 12.54 | 12.62 | 0.0248 | 0.0220 | 0 | 0.0248 | 0.0159 | 0 |
| | 12.82 | 12.73 | 12.77 | 12.88 | 0.0068 | 0.0038 | 0 | 0.0068 | 0.0046 | 0 |
| 15.February.2017 | 14.13 | 13.69 | 13.84 | 14.13 | 0.0204 | 0.0311 | 0 | 0.0204 | 0.0003 | 0 |
| | 14.13 | 14.01 | 14.28 | 14.38 | 0.0087 | 0.0107 | 0 | 0.0087 | 0.0175 | 0 |
| | 14.13 | 13.76 | 14.01 | 14.02 | 0.0265 | 0.0084 | 0 | 0.0265 | 0.0075 | 0 |
| | 14.13 | 14.02 | 14.08 | 14.31 | 0.0077 | 0.0034 | 0 | 0.0077 | 0.0128 | 0 |
| | 12.82 | 13.85 | 13.85 | 14.04 | 0.0801 | 0.0806 | 0 | 0.0801 | 0.0955 | 0 |
| | 12.82 | 13.96 | 14.02 | 14.10 | 0.0886 | 0.0940 | 0 | 0.0886 | 0.0038 | 0 |
| | 12.82 | 13.93 | 14.09 | 14.10 | 0.0863 | 0.0994 | 0 | 0.0863 | 0.0997 | 0 |
| | 12.82 | 12.54 | 12.59 | 12.75 | 0.0220 | 0.0177 | 0 | 0.0220 | 0.0056 | 0 |
| | 12.82 | 12.66 | 12.71 | 12.73 | 0.0122 | 0.0082 | 0 | 0.0122 | 0.0068 | 0 |

|  | 12.82 | 12.65 | 12.67 | 12.81 | 0.0136 | 0.0119 | 0 | 0.0136 | 0.0005 | 0 |
|---|---|---|---|---|---|---|---|---|---|---|
| 23.January.2017 | 15.13 | 14.96 | 14.97 | 15.02 | 0.0112 | 0.0109 | 0 | 0.0112 | 0.0070 | 0 |
| | 15.13 | 14.91 | 14.95 | 14.99 | 0.0143 | 0.0120 | 0 | 0.0143 | 0.0094 | 0 |
| | 15.13 | 14.90 | 14.90 | 14.99 | 0.0151 | 0.0152 | 0 | 0.0151 | 0.0094 | 0 |
| | 13.73 | 11.83 | 12.61 | 12.66 | 0.1383 | 0.0814 | 0 | 0.1383 | 0.0780 | 0 |
| | 13.73 | 11.58 | 12.06 | 12.15 | 0.1564 | 0.1218 | 0 | 0.1564 | 0.0109 | 0 |
| | 13.73 | 11.64 | 11.76 | 11.96 | 0.1525 | 0.1433 | 0 | 0.1525 | 0.1288 | 0 |
| | 13.73 | 11.91 | 11.98 | 12.32 | 0.1326 | 0.1272 | 0 | 0.1326 | 0.1028 | 0 |
| | 13.73 | 11.93 | 12.24 | 12.30 | 0.0113 | 0.1088 | 0 | 0.0113 | 0.1041 | 0 |
| | 13.73 | 11.94 | 12.07 | 12.16 | 0.1307 | 0.1205 | 0 | 0.1307 | 0.1146 | 0 |
| | 13.73 | 11.95 | 11.71 | 11.89 | 0.1296 | 0.1469 | 0 | 0.1296 | 0.1340 | 0 |
| 30.December.2017 | 19.32 | 18.79 | 19.27 | 19.29 | 0.0273 | 0.0028 | 0 | 0.0273 | 0.0018 | 0 |
| | 16.68 | 19.02 | 19.17 | 19.35 | 0.1402 | 0.1495 | 0 | 0.1402 | 0.1599 | 0 |
| | 15.23 | 19.17 | 19.30 | 19.36 | 0.2588 | 0.2671 | 0 | 0.2588 | 0.2712 | 0 |
| | 15.23 | 18.67 | 19.03 | 19.04 | 0.2256 | 0.2495 | 0 | 0.2256 | 0.2505 | 0 |
| | 15.23 | 19.01 | 19.27 | 19.62 | 0.2481 | 0.2652 | 0 | 0.0071 | 0.2880 | 0 |
| | 15.23 | 18.54 | 18.88 | 18.98 | 0.2171 | 0.2397 | 0 | 0.2171 | 0.2462 | 0 |
| | 15.23 | 19.04 | 19.13 | 19.16 | 0.2500 | 0.2563 | 0 | 0.2500 | 0.2583 | 0 |
| | 15.23 | 19.00 | 19.07 | 19.07 | 0.2474 | 0.2521 | 0 | 0.2474 | 0.2524 | 0 |
| | 15.23 | 19.11 | 19.02 | 19.28 | 0.2545 | 0.2485 | 0 | 0.2545 | 0.2659 | 0 |
| | 15.23 | 18.92 | 19.20 | 19.42 | 0.2424 | 0.2603 | 0 | 0.2424 | 0.2751 | 0 |

Russell 2000® Index (RUT)

| | Real Price($) | BS Price | Heston Price | MSV Price | BS Error | Heston Error | Compare | BS Error | MSV Error | Compare |
|---|---|---|---|---|---|---|---|---|---|---|
| 7.March.2017 | 1374.80 | 1367.01 | 1369.98 | 1383.73 | 0.0057 | 0.0035 | 0 | 0.0057 | 0.0065 | 0 |
| | 1374.72 | 1341.12 | 1341.45 | 1341.70 | 0.0244 | 0.0242 | 0 | 0.0244 | 0.0074 | 0 |
| | 1374.72 | 1338.58 | 1339.82 | 1340.91 | 0.0263 | 0.0254 | 0 | 0.0263 | 0.0246 | 0 |
| | 1374.72 | 1340.25 | 1340.38 | 1341.70 | 0.0251 | 0.0250 | 0 | 0.0251 | 0.0240 | 0 |
| | 1374.72 | 1340.13 | 1340.17 | 1340.45 | 0.0252 | 0.0078 | 0 | 0.0252 | 0.0249 | 0 |
| | 1375.03 | 1337.16 | 1339.68 | 1340.21 | 0.0275 | 0.0257 | 0 | 0.0275 | 0.0253 | 0 |
| | 1375.03 | 1338.54 | 1339.68 | 1341.44 | 0.0265 | 0.0257 | 0 | 0.0265 | 0.0244 | 0 |
| | 1375.03 | 1337.61 | 1339.26 | 1339.83 | 0.0272 | 0.0260 | 0 | 0.0272 | 0.0256 | 0 |
| | 1375.03 | 1340.53 | 1341.12 | 1342.11 | 0.0251 | 0.0247 | 0 | 0.0251 | 0.0239 | 0 |
| | 1338.73 | 1339.03 | 1340.65 | | 0.0264 | 0.0262 | 0 | 0.0264 | 0.0250 | 0 |
| 3.March.2017 | 1392.72 | 1385.68 | 1387.63 | 1392.44 | 0.0051 | 0.0037 | 0 | 0.0051 | 0.0002 | 0 |
| | 1392.72 | 1379.68 | 1392.35 | 1393.10 | 0.0094 | 0.0003 | 0 | 0.0094 | 0.0003 | 0 |
| | 1392.70 | 1391.29 | 1391.22 | 1392.93 | 0.0010 | 0.0011 | 0 | 0.0010 | 0.0002 | 0 |
| | 1383.64 | 1384.28 | 1391.94 | | 0.0066 | 0.0062 | 0 | 0.0066 | 0.0007 | 0 |
| | 1392.90 | 1386.67 | 1387.92 | 1388.78 | 0.0103 | 0.0036 | 0 | 0.0103 | 0.0030 | 0 |
| | 1392.90 | 1386.29 | 1389.24 | 1391.49 | 0.0047 | 0.0026 | 0 | 0.0047 | 0.0010 | 0 |
| | 1393.38 | 1305.58 | 1380.94 | 1390.14 | 0.0630 | 0.0089 | 0 | 0.0630 | 0.0023 | 0 |
| | 1393.38 | 1378.92 | 1379.63 | 1390.73 | 0.0104 | 0.0099 | 0 | 0.0104 | 0.0019 | 0 |
| | 1379.57 | 1381.24 | 1390.89 | | 0.0099 | 0.0087 | 0 | 0.0099 | 0.0018 | 0 |
| | 1392.90 | 1381.17 | 1381.30 | 1388.22 | 0.0084 | 0.0083 | 0 | 0.0084 | 0.0034 | 0 |
| 15.Feburary.2017 | 1403.28 | 1399.13 | 1401.71 | 1402.64 | 0.0011 | 0.0030 | 0 | 0.0011 | 0.0005 | 0 |
| | 1403.28 | 1400.27 | 1401.42 | 1401.80 | 0.0021 | 0.0013 | 0 | 0.0021 | 0.0011 | 0 |
| | 1403.28 | 1399.99 | 1401.54 | 1401.60 | 0.0023 | 0.0012 | 0 | 0.0023 | 0.0012 | 0 |
| | 1403.28 | 1399.31 | 1399.09 | 1401.43 | 0.0028 | 0.0030 | 0 | 0.0028 | 0.0013 | 0 |
| | 1405.24 | 1396.16 | 1398.95 | | 0.0063 | 0.0056 | 0 | 0.0063 | 0.0037 | 0 |
| | 1404.08 | 1395.24 | 1396.16 | 1398.95 | 0.0051 | 0.0044 | 0 | 0.0051 | 0.0037 | 0 |
| | 1404.08 | 1396.93 | 1397.97 | 1398.95 | 0.0180 | 0.0132 | 0 | 0.0180 | 0.0130 | 0 |
| | 1403.82 | 1385.47 | 1385.47 | 1385.77 | 0.0111 | 0.0110 | 0 | 0.0111 | 0.0105 | 0 |
| | 1404.03 | 1388.46 | 1388.63 | 1389.22 | 0.0116 | 0.0109 | 0 | 0.0116 | 0.0107 | 0 |
| | 1404.03 | 1387.75 | 1388.78 | 1388.97 | 0.0117 | 0.0108 | 0 | 0.0117 | 0.0086 | 0 |
| | 1404.03 | 1387.67 | 1388.80 | 1391.89 | | | | | | |

## 6. CONCLUSION

The contribution of this paper are threefold. First and the main contribution is that a new random volatility model is used, named as high order moments-based stochastic volatility model (MSV), in which the volatility is a function of time with its level being modulated by a random variable. By using a Taylor series expansion of the option price, it's shown that the model yields an easy formula for approximate option prices and hence can be calibrated extremely fast. The proposed model can even be implemented on a spreadsheet. Secondly, we have demonstrated through comprehensive numerical experiments that MSV model is very competitive in terms of accuracy with Heston model and BS model, while being computationally significantly cheaper to calibrate. Lastly, we have backed up our claims for the usefulness of our model with simulation experiments for comparison of European option prices in all three models. MSV model thus provides a competitive alternative to the existing option pricing models; it is particularly suitable for high frequency financial trading due to its speed of calibration. And as a last note, we conclude that MSV model is more accurate and is the best method for traders to use this model for hedging purpose.